\documentclass[12pt,preprint]{aastex}
\usepackage{natbib}
\usepackage{graphicx}

\begin{document}
\title{A White Light Megaflare on the \lowercase{d}M4.5\lowercase{e} Star YZ CM\lowercase{i}$^1$\footnotetext[1]{\lowercase{\uppercase{B}ased on observations obtained with the \uppercase{A}pache \uppercase{P}oint \uppercase{O}bservatory 3.5-meter telescope, which is owned and operated by the \uppercase{A}strophysical \uppercase{R}esearch \uppercase{C}onsortium.}}
}

\author{Adam F. Kowalski\altaffilmark{2},
        Suzanne L. Hawley\altaffilmark{2},
	Jon A. Holtzman\altaffilmark{3},
        John P. Wisniewski\altaffilmark{2,4},
	Eric J. Hilton\altaffilmark{2}
       }

\altaffiltext{2}{Astronomy Department, University of Washington,
   Box 351580, Seattle, WA  98195 \\
email: kowalski@astro.washington.edu}
\altaffiltext{3}{Department of Astronomy, New Mexico State University, Box 30001, Las Cruces, NM 88003}
\altaffiltext{4}{NSF Astronomy \& Astrophysics Postdoctoral Fellow}

\begin{abstract}

On UT 2009 January 16, we observed a white light megaflare on the dM4.5e star YZ CMi as part of a long-term spectroscopic flare-monitoring campaign to constrain the spectral shape of optical flare continuum emission.  Simultaneous $U$-band photometric and 3350-9260\AA$ $ spectroscopic observations were obtained during 1.3 hours of the flare decay.  The event persisted for more than 7 hours and at flare peak, the $U$-band flux was almost 6 magnitudes brighter than in the quiescent state.  The properties of this flare mark it as one of the most energetic and longest-lasting white light flares ever to be observed on an isolated low-mass star.  We present the $U$-band flare energetics and a flare continuum analysis.  For the first time, we show convincingly with spectra that the shape of the blue continuum from 3350\AA$ $ to 4800\AA$ $ can be represented as a sum of two components: a Balmer continuum as predicted by the Allred et al radiative hydrodynamic flare models and a $T\sim$10,000K blackbody emission component as suggested by many previous studies of the broadband colors and spectral distributions of flares.  The areal coverage of the Balmer continuum and blackbody emission regions vary during the flare decay, with the Balmer continuum emitting region always being significantly ($\sim$3-16 times) larger. These data will provide critical constraints for understanding the physics underlying the mysterious blue continuum radiation in stellar flares.

\end{abstract}


\section{Introduction}

Due to their strong and persistent surface magnetic fields, active M dwarfs can produce frequent flares, sometimes lasting for many hours and reaching luminosities that approach a significant fraction of the star's bolometric luminosity \citep{Kunkel1969, Moffett1974, Bond1976, Hawley1991, Favata2000, Robinson2005, Osten2007, Kowalski2009}.  During both solar and stellar flares, emission is seen as line and continuum radiation from X-ray to radio wavelengths, with a prominent mode of radiative energy release occurring in the blue and near-ultraviolet (NUV) continuum, which is commonly referred to as the white light continuum.  Unfortunately, the spectral components which comprise the white light continuum are not well-constrained, and therefore the physics of the underlying flare mechanism that gives rise to such a conspicuous signature remains a mystery.

Significant effort has gone into characterizing the white light continuum in stellar flares, beginning with a single component (hydrogen recombination) model \citep{Kunkel1969, Kunkel1970}.  
A two-component spectral model consisting of hydrogen recombination and an impulsively heated photosphere was first proposed by \cite{Kunkel1970} as a possible explanation of the spread in broadband flare colors.  Time-resolved multi-channel photometry \citep{Mochnaki1980} and spectrometry \citep{Kahler1982} were obtained longward of 4200\AA$ $ for flares on YZ CMi and UV Ceti;  at peak, these flare spectra were fit with a Planck function of $T=7400-9500$K, thereby showing evidence of a hot optically thick component present during the impulsive phases of flares.  Instrumental effects shortward of the Balmer jump (3646\AA) prevented a precise characterization of the Balmer continuum in the data of \cite{Mochnaki1980}.

 Higher resolution spectra in the blue/NUV were obtained during large flares on AD Leo \citep[3560\AA - 4440\AA;][]{Hawley1991} and CN Leo \citep[3050\AA - 3860\AA;][]{Fuhrmeister2008} showing a nearly continuous rise into the NUV with a $T\sim8400-11,300$K blackbody; surprisingly no discontinuity at the Balmer jump was seen in either of these events \citep[see also ][]{Eason1992, Garcia2002}.  The white light continuum has therefore been attributed to a single dominant blackbody component with $T\sim9000-10,000$K \citep[e.g.,][]{Hawley2003}.  In contrast, flare models predict spectra that are dominated by hydrogen bound-free continua, which
provide a poor match to the observations \citep[ hereafter A06]{Hawley1992, Allred2006}.

A multi-component model was reintroduced by \cite{Zhilyaev2007}, who used high-cadence $UBVRI$ photometric observations of the dM3.5e star EV Lac to suggest that the white light emission consists primarily of blackbody radiation at flare peak and hydrogen continuum during the flare decay.  However, significant ambiguity results when using broadband photometry to characterize the white light continuum.  A06 showed that a flare model spectrum that included a large Balmer jump, when convolved with broadband $UV+UBVR$ filters, actually resembles the shape of a $T\sim$9,000K blackbody.  To unambiguously identify the spectral components present during stellar flares, we have therefore begun to compile a large catalog of time-resolved optical/NUV flare spectra on several nearby flare stars.  In this Letter, 
we present time-resolved spectra obtained during a large flare on the dM4.5e star YZ CMi and use results from the recent radiative hydrodynamic (RHD) flare models of A06
to show that both Balmer continuum and hot blackbody components are present in the white light continuum emission.  An extensive paper describing our flare catalog, including discussion of the flare emission lines and comparison with models will be forthcoming.

\section{Observations}
\subsection{NMSU 1-m Photometry}
Photometric observations were obtained in the Johnson $U$ band on UT 2009 January 16 with the New Mexico State University 1-m Telescope at the Apache Point Observatory (APO).  The 1-m is operated robotically and is open for observations when the ARC 3.5-m Telescope, located at the same site, is open \citep{Holtzman2010}.  The dM4.5e star YZ CMi was observed for nearly 8 hours with an exposure time of 10s; with readout the photometric cadence ranged between 20 - 27s.  We performed differential photometry on YZ CMi using the comparison star HD 62525. The night was clear and photometric based on the data from nearby comparison stars.  The $U$-band light curve of YZ CMi is shown in Figure \ref{fig:figs}a.

\subsection{APO 3.5-m Spectroscopy}
We obtained NUV/optical spectroscopy for 1.3 hours during the decay phase of the flare, denoted by vertical lines in Figure \ref{fig:figs}a, using the Dual Imaging Spectrograph (DIS) with the ARC 3.5-m Telescope at APO.  We used the B400$+$R300 gratings, providing wavelength coverage from 3350\AA$-$9260\AA.  The observations were taken through a 1.5 $\arcsec$ slit which yielded resolutions of $R\sim625$ (at 4000\AA) and $R\sim980$ (at 6563\AA), determined from the FWHM of the arc lamp lines.  The slit was oriented at the parallactic angle in order to compensate for atmospheric differential refraction \citep[see ][]{Filippenko1982}.  We observed YZ CMi with exposure times of 10s, and with readout the total cadence was $\sim$28s.  In total, 163 flare spectra were obtained simultaneously in both the blue and red bandpasses.
The data were reduced using standard IRAF\footnote{IRAF is distributed by the National Optical 
Astronomy Observatories, which are operated by the Association of Universities for Research in 
Astronomy, Inc., under cooperative agreement with the National Science Foundation} procedures, wavelength calibrated using 
HeNeArHg lamp spectra, and flux-calibrated by comparing observations of the spectrophotometric standard star Feige 34 to the accepted fluxes of \cite{Oke1990}.  To correct for variable grey slit loss (e.g., due to guiding errors), we also scaled each spectrum during the flare using the $U$-band light curve.  A spectrum of YZ CMi in its quiescent (non-flaring) state was obtained with the same instrumental setting on UT 2008 November 24 and was used for comparison (see \S 3.2).\footnote{The quiescent spectrum was convolved with the Johnson $B$ and $V$ bands (using filter curves from \cite{MA2006}) and scaled to match the flux from the literature \citep[$V = 11.2$, $B = 12.8$;][]{Reid2005}}

\section{Analysis}
\subsection{$U$-Band Flare Energetics}
   This flare event is remarkable in its total energy, peak luminosity, total duration, and emission morphology.  A small precursor, sometimes seen prior to large flares \citep{Moffett1974}, began at 04:14:54UTC which increased the $U$-band flux by a factor of $\sim$2, and represents the onset of the flare event.  At 04:32:00UTC, there was a $\sim$10 min impulsive rise to flare peak, where the $U$-band flux was a factor of $\sim$200 times the quiescent value.  The decay phase included many smaller events, and the $U$-band flux was still elevated by over a magnitude above quiescence at the end of the night.  In \S 4, we discuss the extremely long duration of this event.
   
  The total $U$-band flare energy is used to quantify the energy release in continuum radiation during stellar flares \citep{Moffett1974}.  We integrate under the light curve to obtain an equivalent duration \citep{Gershberg1972} of $4.25 \times 10^{5}$s, which is the amount of time that the star would spend emitting at the pre-flare level to produce the same total amount of energy as the flare.   Multiplying the equivalent duration by the non-flaring $U$-band luminosity of 4.00 $\times 10^{28}$ ergs s$^{-1}$ \citep{Moffett1974} gives $E_{U,flare} > 1.7 \times 10^{34}$ ergs (this is a lower limit since flare emission is still seen at the time the light curve ends).  At peak, the flare was emitting with a $U$-band luminosity of 8.3$\times 10^{30}$ ergs s$^{-1}$, or $\sim$37\% the stellar bolometric luminosity \citep[M$_{Bol}$= 10.25,][]{Reid2005}.  
This flare was $\sim$100 times more energetic and persisted for more than 10 times as long as the largest $U$-band flare event observed on this star during 55 hours of monitoring from the large statistical flare study of \cite{Moffett1974} and \cite{Lacy1976}.  Although it is unknown if the power law fit to the flare-frequency distribution of \cite{Lacy1976} holds at high flare energies, an extrapolation predicts that flares with $U$-band energies $> 1.7 \times 10^{34}$ ergs occur at a rate of $\sim$once per month.

\subsection{The Flare Continuum}
 
\begin{figure}[htbp]
\centering
 \includegraphics[scale=0.3,angle=90]{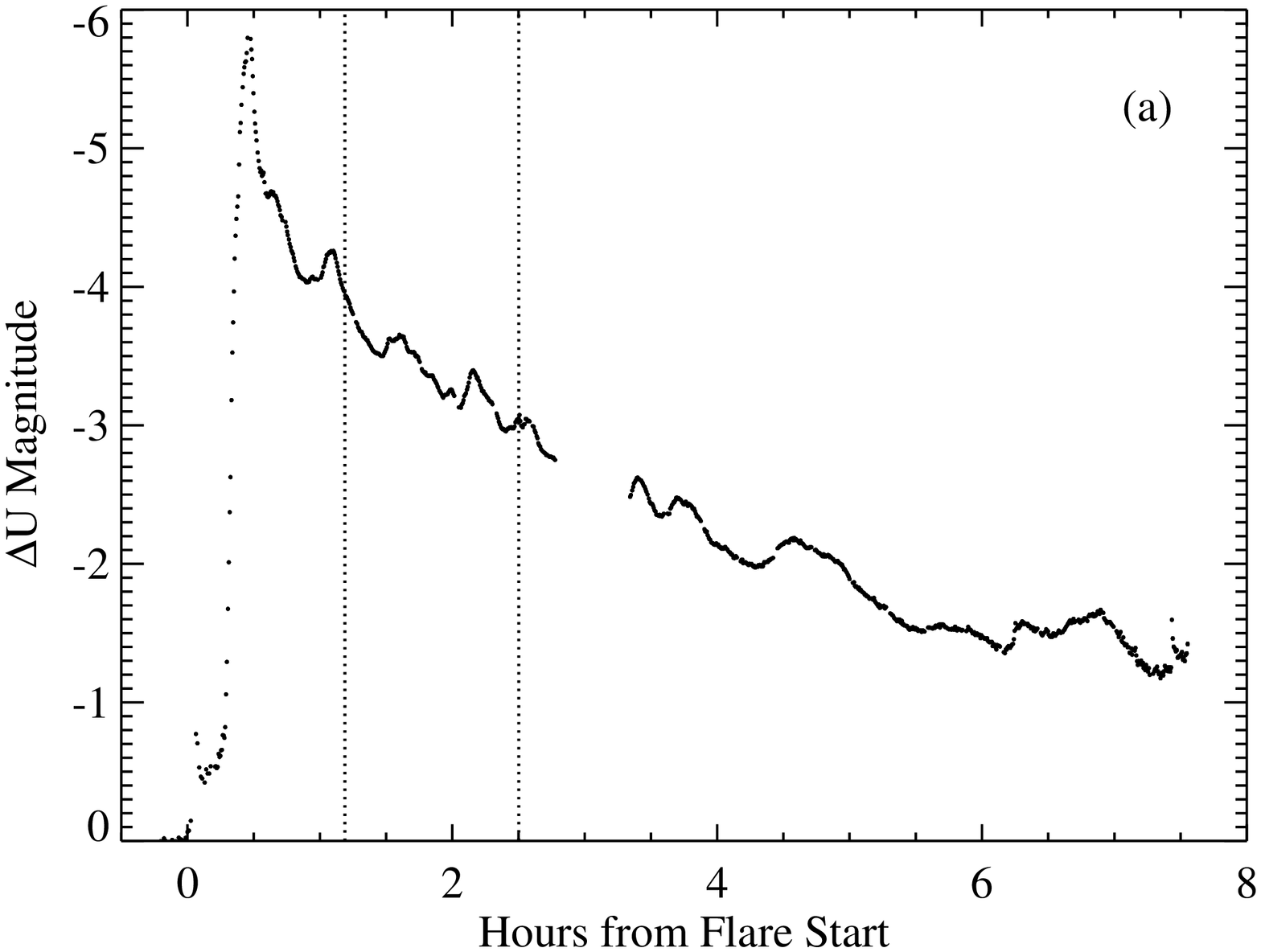} 
  \includegraphics[scale=0.3,angle=90]{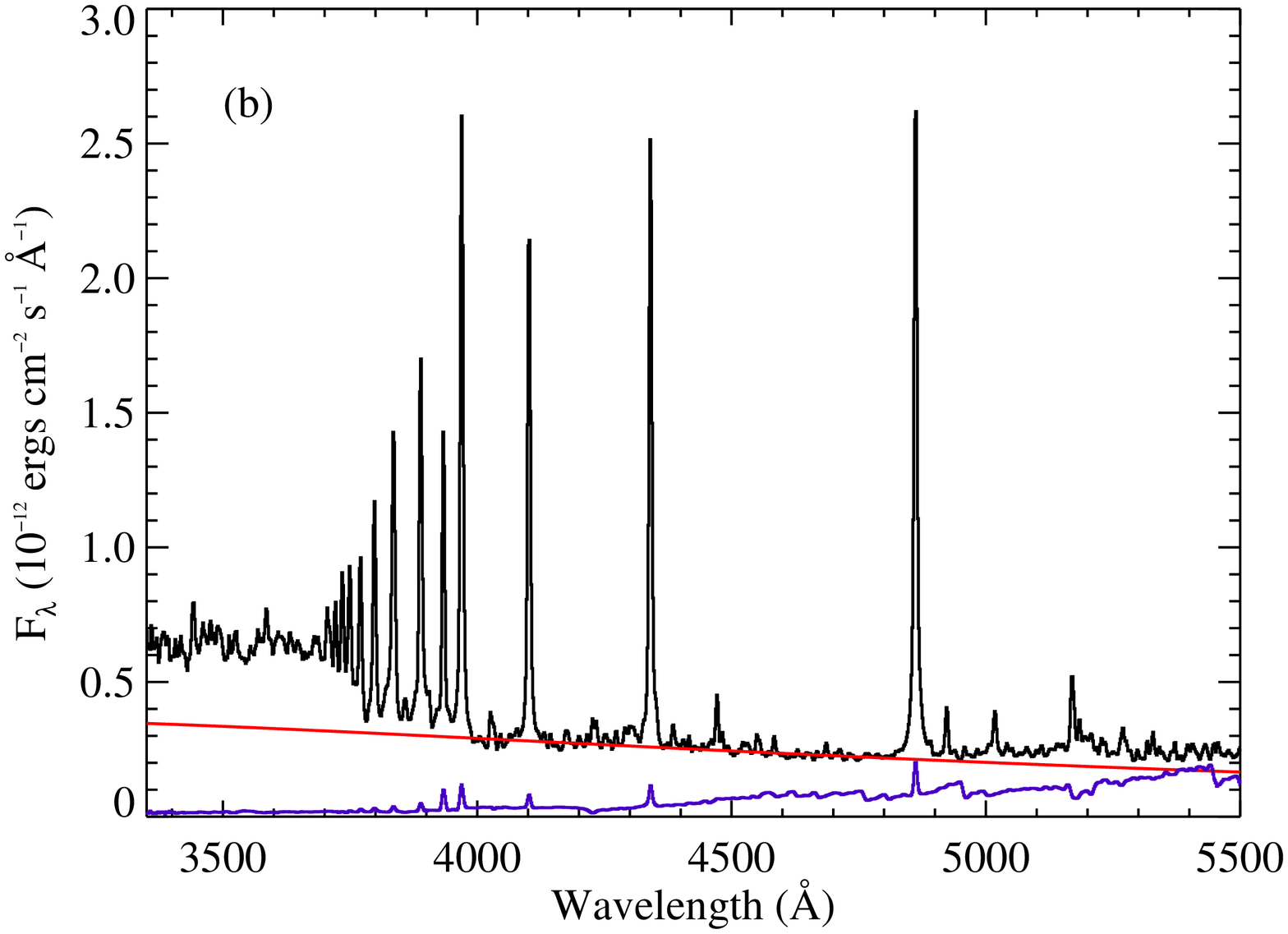}
   \includegraphics[scale=0.3,angle=90]{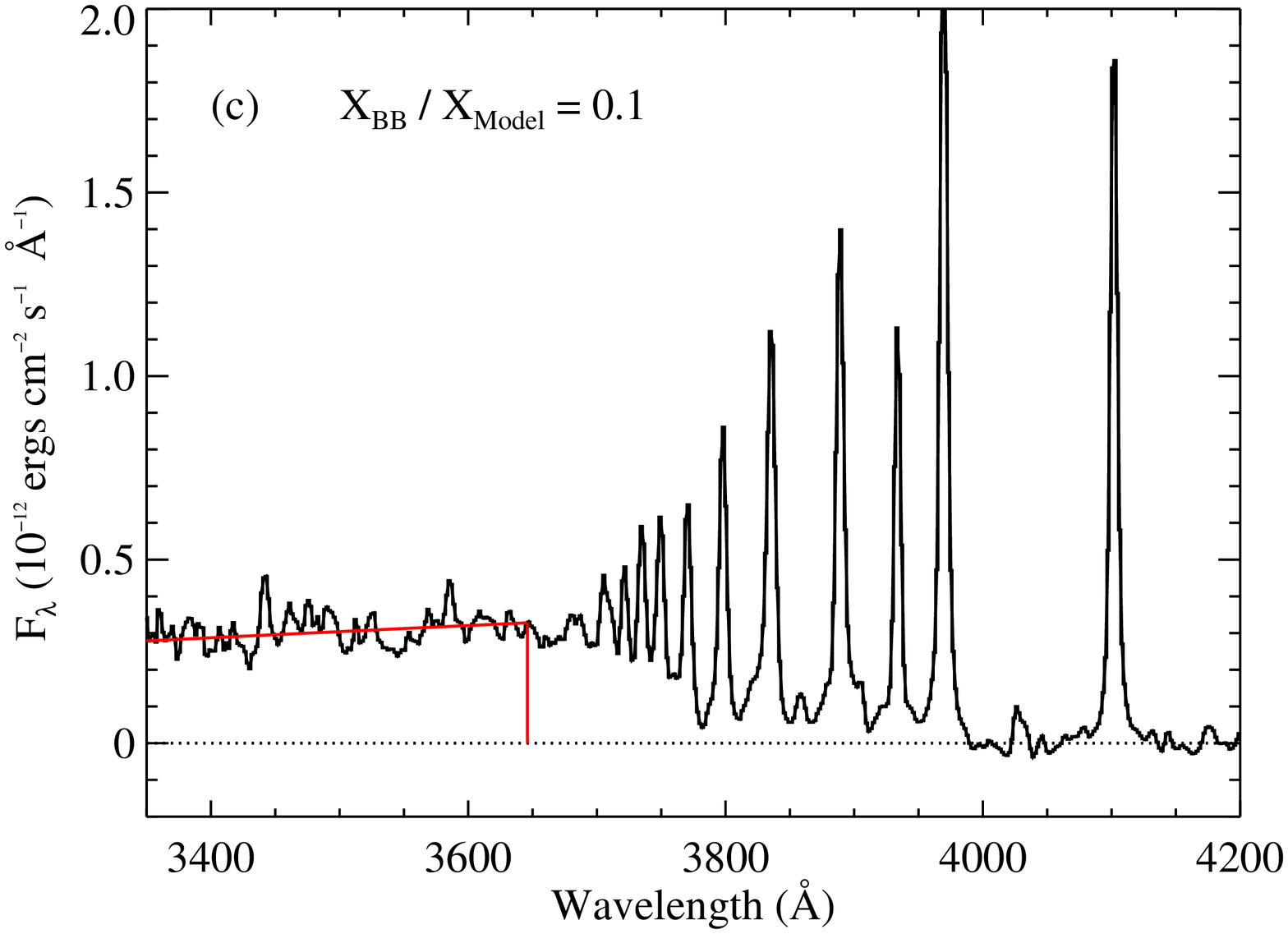}
   \includegraphics[scale=0.3,angle=90]{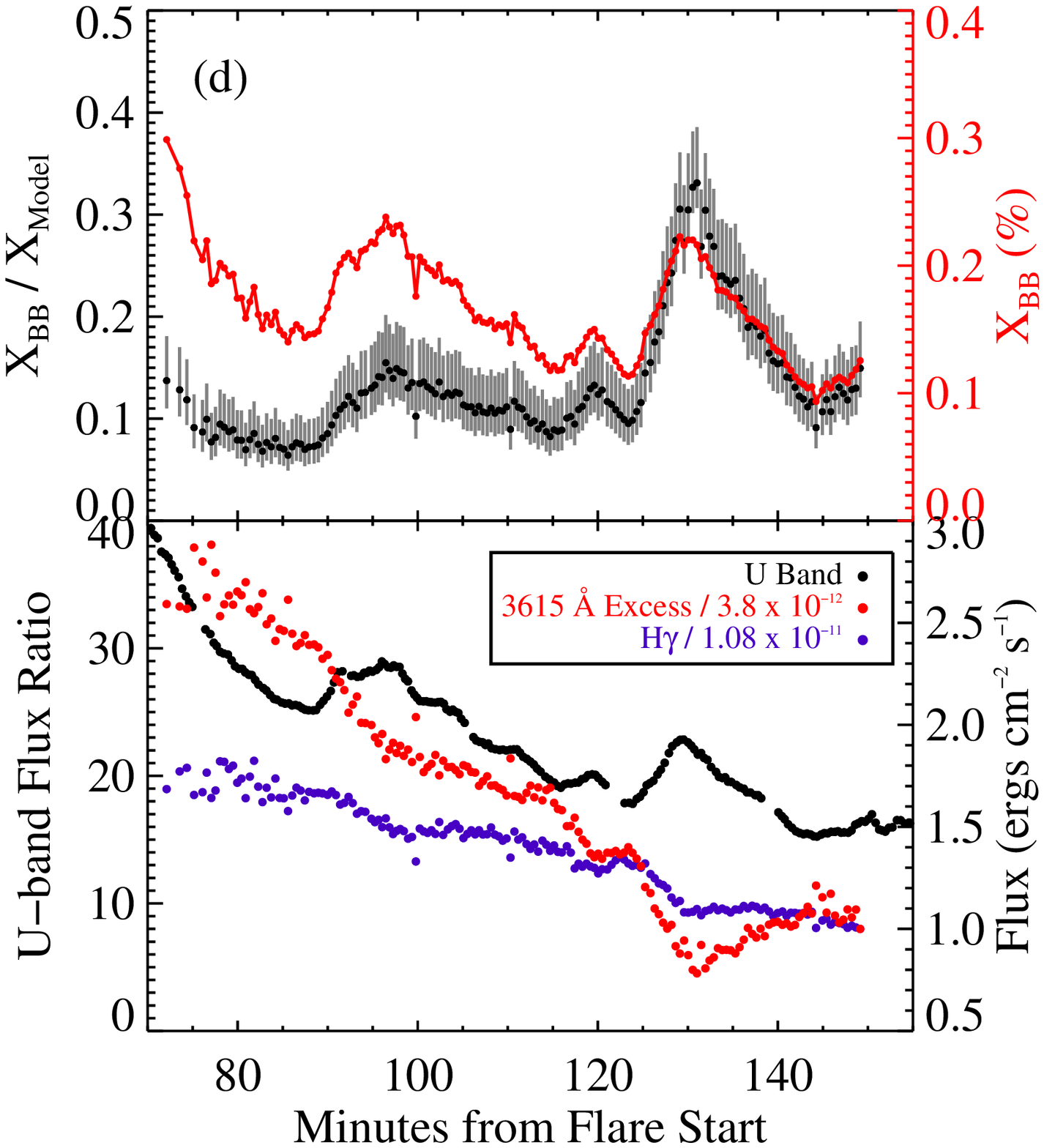}
    \caption{(a) The flare $U$-band light curve with vertical lines indicating the period of spectroscopic observations. (b) A flare spectrum at $t=76.6$min after flare start.  The quiescent spectrum is shown in purple for comparison.  The continuum from 4000 - 4800\AA$ $  was fit with a $T=10,000$K blackbody and a filling factor of 0.22\%.  There is excess emission above the blackbody blueward of 3800\AA.  (c) The 10,000K blackbody component has been subtracted from the spectrum in panel (b) and the F11 flare spectrum from A06 fits the excess (Balmer) continuum at wavelengths shorter than 3646\AA.
(d) (upper panel) The filling factor ratio ($X_{BB}/X_{Model}$, black points) and the inferred area coverage of the blackbody component (red points).  (lower panel)  The time-evolution of the $U$-band (left axis), and the Balmer continuum and H$\gamma$ fluxes (right axis). }
\label{fig:figs}
\end{figure}

  The spectral data were obtained while the $U$-band emission was still elevated 15-37 times (3-4 mag) over quiescence.  Complex morphology is seen in the $U$-band light curve during this time period (see Figure \ref{fig:figs}a), allowing us to study the flare continuum variations during several smaller impulsive events. Figure \ref{fig:figs}b shows a spectrum ($t=76.6$min after the flare start) from 3350\AA$-$5500\AA$ $ near the beginning of our spectroscopic observations during the $U$-band decay from a large secondary emission peak.  A quiescent spectrum of YZ CMi (see \S2.2) is also shown in Figure \ref{fig:figs}b and has been subtracted from the flare spectrum to give a ``flare-only'' spectrum.  Our spectra have the highest time resolution with high signal to noise in the blue/NUV (S/N $>$ 50 at 3600\AA) ever obtained during
a large stellar flare.  They are also unique in having simultaneous high quality photometric information so that we may relate the spectral evolution to the light curve morphology.  

 We have for the first time convincingly identified two distinct continuum components that simultaneously contribute to the optical/NUV flux in flare spectra.  The first component dominates the spectrum from 4000-4800\AA$ $ and exhibits the rising trend of a $T\sim$10,000K blackbody, which is plotted on the spectrum in Figure \ref{fig:figs}b using a filling factor, X$_{BB}$, of 0.22\% of the projected stellar disk area.  We follow \cite{Hawley2003} to constrain the filling factor of blackbody emission as a function of time using the equation:

\begin{equation}
F_{fl,\lambda}(t) = X_{BB}(t) \frac{R_{\star}^{2}}{d^2} \pi B_{\lambda}(T_{fl}) 
\end{equation}
where $F_{fl,\lambda}$ is the flare flux observed at Earth in the continuum windows from 3995-4020\AA, 4140-4210\AA, 4425-4450\AA, and 4600-4800\AA;  $R_{\star}=0.3 \times R_{Sun}$,  $d=5.97$$ $ pc \citep{Reid2005} and $T_{fl}=10,000$K.

  From Figure \ref{fig:figs}b, it is clear that the blackbody fit does not explain all of the continuum emission at wavelengths
shorter than $\sim$3800\AA.  Starting near H10 (3799\AA), there is a rise to H15$+$H16, the last two apparent Balmer lines in emission.  Blueward of $\sim$3700\AA, the flare spectrum becomes flat into the NUV, and there is excess flux visible 
above the 10,000K blackbody at these wavelengths.

A06 presented RHD flare simulations for a medium (F10) and
a strong (F11) impulsive flux of nonthermal electrons into an M dwarf 
atmosphere.  The A06 models predict strong Balmer continuum radiation with a 
noticeable Balmer jump at 3646\AA, together with a photosphere that is warmed
by $\sim$1200K to a temperature of 4600K.  We isolated the Balmer continuum
component of the A06 F11 model spectrum taken at the last time-step of
their simulation ($t=15.9$s), and fit it to the observed flare spectra.
Figure \ref{fig:figs}c shows the flare spectrum 
from Figure \ref{fig:figs}b with the 10,000K blackbody component subtracted and
the model Balmer continuum component scaled to the excess emission in the 
3600\AA\ $ < \lambda < 3646$\AA\ region.  The shape of the observed excess emission is well-described by the F11 flare model prediction (shown in red).  Although there is no abrupt discontinuity indicative of a Balmer jump in our observed spectrum, 
we posit that blending \citep[possibly due to Stark broadening; ][]{Donati1985} 
of the higher order Balmer lines in the 3646\AA\ $< \lambda <$ 3800\AA\ region 
effectively hides the sharp discontinuity that is seen in the model spectrum.

\subsection{Relative Filling Factors}
The filling factor for the F11 model spectrum is calculated by scaling the model 
flux in the 3600-3646\AA\ region ($F_{\lambda, Model}=1.14 \times 10^{7}$ ergs cm$^{-2}$ s$^{-1}$ \AA$^{-1}$) to the flux excess observed at Earth after 
subtracting the 10,000K blackbody. The contribution function for the 
Balmer continuum in the F11 model shows that the emission arises from a relatively thin layer in the atmosphere (J. Allred, private communication); it is therefore reasonable to approximate changes in the filling factor to changes in the emitting area of the Balmer continuum.

In the upper panel of Figure \ref{fig:figs}d (black points), we show 
the ratio of the filling factors for a 10,000K blackbody and the
model Balmer continuum component.  The error bars are obtained by using 9000 and 11,000K temperatures for the blackbody.  The ratio for the sample spectrum in Figure \ref{fig:figs}b is $X_{BB}/X_{Model}=0.10 \pm ^{0.04}_{0.02}$. The filling
factor ratio indicates that the two emission components exhibit significantly 
different time evolution, with the blackbody source emitting from a region 
$\sim$3-16 times smaller than the source of the Balmer continuum.  

The inferred area filling factor ($X_{BB}$) of the blackbody component 
is shown separately (red points) and exhibits very similar morphology to 
the $U$-band light curve (reproduced in the lower panel). Although we expect 
the \emph{total} continuum flux to follow the $U$-band since we scaled the 
spectra to achieve absolute flux calibration (\S3.2), it is surprising 
that \emph{only} the blackbody component follows the $U$ band.\footnote{These trends are also visible in the data before the flux calibration scaling was applied.}  It appears that the deviations from the general flare decay seen 
in the $U$ band primarily arise from the changing area of the blackbody 
emission region.  When the deviations are strongest near the peaks at $t\sim95$ min and $t\sim130$ min, the spectra are increasingly dominated by the blackbody component.  This is consistent with the flare-peak spectra from \cite{Kahler1982}, \cite{Hawley1991}, and \cite{Fuhrmeister2008} and the photometry of \cite{Zhilyaev2007}.  We also note preliminary evidence of an apparent increase in the best-fit blackbody temperature to $T\sim13,000$K at the peak near $t\sim130$ min.

The derived filling factor ratios should be regarded as illustrative 
since our Balmer continuum spectrum from the F11 model of A06
assumes that the 
nonthermal electron flux is constant at 10$^{11}$$ $ ergs cm$^{-2}$ s$^{-1}$.  
Also, we use the model flare spectrum from only one time-step in the RHD 
simulation.  It is likely that the nonthermal electron flux varied both
with time and over the geometry of the flaring region.  Any further 
assessment is beyond the scope of this Letter, but we note that the less 
energetic A06 F10 flare spectrum at $t=230.0$s gives 
nearly 10 times larger filling factors for the Balmer continuum.

\subsection{Balmer Line and Continuum Radiation}

The time-evolution of the Balmer continuum closely follows the hydrogen 
Balmer lines.  We show the variation in the Balmer continuum (using the integrated
flux in a 30\AA\ window centered at 3615\AA) compared to the H$\gamma$ line flux 
and the $U$ band in the lower panel of Figure \ref{fig:figs}d. The fluxes are normalized to the last spectral observation we obtained. 
The Balmer continuum shows a slow decay throughout our observations, as is typical of the Balmer line radiation \citep{Hawley1991}.  This overall decay is probably extended emission from the previous, much more luminous, impulsive events
seen in the $U$-band light curve.  The larger variations in the Balmer 
continuum compared to the H$\gamma$ emission may be partially due to 
small errors in isolating the Balmer continuum by subtracting the 
underlying blackbody continuum (the H$\gamma$ flux is well measured as 
it has a local continuum).  As noted above in the filling factor discussion and evident by comparing the upper and lower panels of Figure \ref{fig:figs}d, 
the Balmer line and continuum 
radiation exhibit quite different time evolution compared to the blackbody emission
component.  We propose two possible explanations: 

\begin{description}
\item[(a)] The hydrogen-emitting regions exhibit a delayed and more 
gradual response to each successive impulsive (blackbody) event. 
In this scenario, the gradual increase in hydrogen Balmer emission around $t=122$ and $t = 137$ min are associated with the impulsive events in blackbody emission (upper panel) at $t=117$ min and $t=127$ min, respectively.  Also, Balmer emission around $t=80$ min and $t=105$ min is elevated yet only slowly decreasing, and could be gradually declining emission from the large $U$-band peaks at $t\sim60$ min and $t\sim95$ min, respectively.

\item[(b)] The time evolution of the blackbody emission and the Balmer emission may be anti-correlated.  If true, this may provide an important constraint on the origin of the blackbody emission.  The possible anti-correlation is most apparent during the $U$-band rise starting at $t=124$ min.  In $\sim5$ min the Balmer continuum decreases by $\sim$40\% while the blackbody filling factor almost doubles. Hydrogen continuum dimming is predicted by flare models \citep{Abbett1999, Allred2005}, but only for very short durations (0.1s) at the onset of a flare.
\end{description}

More detailed models will be required to understand these complex variations, and in particular, the seemingly anti-correlated trends between the blackbody and Balmer continuum and lines during the time intervals when the blackbody filling factor is increasing in Figure \ref{fig:figs}d.

\section{Summary and Speculation}

We observed a white light megaflare on the dM4.5e star YZ CMi in the $U$ band and with simultaneous optical/NUV spectroscopy.  The $U$-band energetics and light curve morphology qualify this flare as an extraordinary and rare event, similar to the $\sim6.6 \times 10^{34}$ erg flare on YZ CMi seen by \cite{Andrews1969}, \cite{Lovell1969}, and \cite{Kunkel1969}.  
Following a solar analogy, we speculate that the flaring region on YZ CMi was a complex arcade of sequentially reconnecting magnetic loops.  Each reconnecting loop 
accelerated a beam of nonthermal electrons that impacted the lower atmosphere,
producing the observed blue/NUV line and continuum emission. The sum of a large
number of individual emitting regions may have enabled this flare to persist 
for such an unusually long time.

Using high time resolution, high signal-to-noise spectra, we have shown 
that the blue/NUV flare continuum radiation can be explained as a 
sum of a $T\sim$10,000K blackbody component and a Balmer continuum component, 
with only the blackbody emission present from 4000-4800\AA.  The relative filling factors of the two components indicate that the Balmer continuum comes from 
a larger region, plausibly originating from the flaring loops at chromospheric heights, where the Balmer lines originate.  This is consistent with the height of the Balmer continuum emission derived from solar flare data \citep{Hudson2010}, and also with the height of formation predicted by the RHD model of A06.  

We found that the blackbody emission arises from a region $\sim$3-16 times 
smaller in area than the Balmer continuum emission region. This blackbody 
emission may possibly originate in 
concentrated magnetic footpoint regions in the lower atmosphere, 
similar to the localized areas that emit in white light 
during solar flares \citep{Metcalf2003, Fletcher2007, Isobe2007, Jess2008},
which are often spatially and temporally coincident 
with impulsive heating by nonthermal electrons inferred from hard X-ray 
observations \citep[e.g.,][]{Rust1975, Hudson1992, Neidig1993, Fletcher2007}.  
In accordance with this scenario, a large complex of photospheric hot spots 
may have been created during the first impulsive events in the YZ CMi 
$U$-band light curve.  As the nonthermal electron beams weakened 
(in energy and/or in spatial extent) so did the areal coverage of 
these hot spots.  This gradual decay extends into the time covered by 
our spectral observations and is seen in the overall decline of the
blackbody areal coverage in the upper panel of Figure \ref{fig:figs}d.  
During our spectral observations, a few new hot spots may have been created 
at the footpoints of newly reconnected loops, causing the transient 
increases seen in the $U$-band and in the effective area of the blackbody 
emitting region.  

We see strong evidence through all of our observations 
for a blackbody continuum emission component with an approximate temperature 
of $10,000$K.  The persistence of the hot blackbody emission indicates 
that a continual source of particle acceleration and plasma heating likely 
still exists during the decay phase of the flare.  Yet, the A06 
models, which employ a nonthermal electron beam as is seen on the Sun, 
predict the photosphere of an M dwarf to be heated by at most $\sim$1200K 
during a large flare.  The physical mechanism which generates the strong 
blackbody emission (presumably from a more strongly heated photosphere) 
therefore remains unknown.  The possible anti-correlation between
the Balmer continuum and blackbody emission components may provide an
important clue to the nature of the heating mechanism that is responsible
for the blackbody emission.

\section{Acknowledgments}
   AFK, SLH, and EJH acknowledge support from NSF grant AST 08-07205.  JPW aknowledges support from NSF AAPF AST 08-02230.  We would like to thank B. Williams, J. Allred, and E. B\"ohm-Vitense for their helpful conversations.

\clearpage

\end{document}